\documentclass[10pt,letterpaper]{article}
\usepackage[margin=0.75in,top=0.72in,bottom=0.75in]{geometry}
\usepackage[utf8]{inputenc}
\usepackage{times}
\usepackage{microtype}
\usepackage{amsmath}
\usepackage{amssymb}
\usepackage{graphicx}
\graphicspath{{../standalone_figures/}{standalone_figures/}}
\usepackage{booktabs}
\usepackage[font=small,labelfont=bf]{caption}
\usepackage{array}
\usepackage{tabularx}
\usepackage{multirow}
\usepackage{xcolor}
\definecolor{alethiacite}{HTML}{001472}
\usepackage{textcomp}
\usepackage[round,authoryear]{natbib}
\let\cite\citep
\usepackage{url}
\usepackage{xurl}
\usepackage[colorlinks=true,linkcolor=black,citecolor=alethiacite,urlcolor=alethiacite]{hyperref}
\usepackage{flafter}
\usepackage{placeins}
\usepackage{float}
\newcommand{\papertitle}{Benchmarking Neural Defend ARCAS 1B: A Foundational Multimodal Deepfake Detection Model}
\newcommand{\paperauthor}{Sivashankar Selvarajan, Piyush Verma, Sumit Kumar, and Sharayu N. Deshmukh}

\newcommand{\searchcutoff}{15 September 2026}

\makeatletter
\renewcommand\section{\@startsection{section}{1}{\z@}%
  {-1.5ex \@plus -0.5ex \@minus -.2ex}%
  {0.8ex \@plus .2ex}%
  {\raggedright\normalfont\large\bfseries}}
\renewcommand\subsection{\@startsection{subsection}{2}{\z@}%
  {-1.25ex\@plus -0.4ex \@minus -.2ex}%
  {0.55ex \@plus .15ex}%
  {\raggedright\normalfont\normalsize\bfseries}}
\renewcommand\subsubsection{\@startsection{subsubsection}{3}{\z@}%
  {-1ex\@plus -0.35ex \@minus -.2ex}%
  {0.4ex \@plus .1ex}%
  {\raggedright\normalfont\normalsize\itshape}}
\makeatother
\renewenvironment{abstract}{\section*{\centering Abstract}\small}{\par}
\newcommand{\IEEEPARstart}[2]{#1#2}

\hypersetup{
  pdftitle={\papertitle},
  pdfauthor={\paperauthor},
  pdfsubject={Image-only benchmark evaluation of Neural Defend ARCAS 1B},
  pdfkeywords={AI-generated image detection, deepfake detection, image forensics, cross-benchmark evaluation}
}
\begin{document}
\twocolumn[
\begin{center}
\vspace*{0.55in}
\rule{\textwidth}{0.8pt}\par
\vspace{0.32in}
{\LARGE\bfseries \papertitle\par}
\vspace{0.28in}
\rule{\textwidth}{0.8pt}\par
\vspace{0.32in}
{\normalsize\bfseries Sivashankar Selvarajan\textsuperscript{1}\quad Piyush Verma\textsuperscript{1}\quad Sumit Singh\textsuperscript{1}\quad Sharayu N.\ Deshmukh\textsuperscript{1}\par}
\end{center}
\vskip 0.28in
]
\footnotetext[1]{\textsuperscript{1}Neural Defend Inc.. Correspondence to: Sivashankar Selvarajan \texttt{<siva@neuraldefend.com>}.}

\begin{abstract}
AI-generated imagery evolves faster than benchmark-specific detector evaluations, making a single score an incomplete account of generalization. This paper evaluates Neural Defend ARCAS 1B across benchmark families without benchmark-specific parameter updates. We retain native aggregation and supplement it with record-level measures, coverage accounting, and subgroup diagnostics. Each Results subsection identifies the release and evaluation population, reports the official metric, and describes observed error patterns. A combined analysis synthesizes shared patterns while preserving the distinction between native and pooled quantities. Cross-paper comparisons are restricted to aligned evidence; differences in release, population, preprocessing, training, or benchmark exposure are context rather than rank. The findings characterize performance on evaluated records, not universal reliability, calibration, attribution, or future adaptive attacks. By keeping benchmark-native outcomes distinct from pooled summaries, the study makes test-population, class-balance, and missing-record-coverage differences visible. It supports interpretation of detector results in research, platform-safety, and forensic-review settings, foregrounding traceable protocol conditions over claims or leaderboard comparisons.
\end{abstract}

\section{Introduction}

\IEEEPARstart AI-generated images span GAN synthesis, diffusion generation, face manipulation, image editing, and online content. Their rapid improvement creates a provenance problem for platforms, investigators, and researchers: detectors face generators, image sources, and post-processing pipelines unseen in development. Early work found shared generator artifacts support transfer \cite{wang2020cnn}, but later work showed transfer can weaken under newer generators, diffusion models, and distribution shifts \cite{ojha2023universal,wang2023dire,tan2023gradients}. Contemporary methods explore training-data diversity, reconstruction cues, frequency information, patch-level texture, and self-supervised representations \cite{yan2025sanity,liu2024fatformer,li2025safe,zhong2026sdaie}. These approaches improve capability but do not make a benchmark score a complete description of detector behavior.

The difficulty is partly methodological. A binary detector can appear strong in overall accuracy while missing a meaningful share of generated images, or it can rank examples well while performing less well at a fixed threshold. Common reported metrics such as accuracy, balanced accuracy, average precision, and AUROC serve different purposes and depend on the test population and aggregation rule. Evaluation releases also differ: Chameleon emphasizes human-curated high-resolution imagery \cite{yan2025sanity}; Community Forensics reports generator-level means \cite{park2025community}; AIGIBench includes generation and manipulation settings \cite{li2025aigibench}; and AIGCDetectBenchmark, WildFake, and GenImage expose different generator families, availability conditions, and evaluation conventions \cite{aigcdetect_evaluation,hong2025wildfake,zhu2023genimage}. An evidence-based benchmark paper must therefore preserve these distinctions rather than converting all results into one apparent league table.

Recent benchmark studies strengthen this requirement. OpenFake distinguishes held-out and naturally circulated splits \cite{livernoche2025openfake,openfake_dataset}; WildFake documents a hierarchical test index with source availability considerations \cite{hong2025wildfake,wildfake_repository}; and GenImage is commonly evaluated through held-out generator folders \cite{zhu2023genimage,genimage_repository}. At a broader scale, a recent multi-detector study reports substantial dataset dependence for out-of-the-box AI-image detection \cite{ren2026outofbox}. Together, these studies motivate a protocol-aware evaluation practice: an unweighted generator mean, split accuracy, available-case accuracy, and threshold-free ranking metric should be interpreted in their native setting. Comparisons with published detectors or commercial systems are informative only when release, population, metric, preprocessing, and benchmark exposure are sufficiently aligned.

This paper evaluates supplied scores from the Neural Defend ARCAS 1B image detector on the available records from seven selected public benchmark families. Neural Defend specifically attests that the detector was frozen and received no reported benchmark-specific parameter updates; the supplied package does not independently establish those claims. Each dedicated Results subsection states the benchmark release and denominator, retains the official metric as its headline result, examines error patterns, and reviews available baseline, detector, research-paper, or company evidence for that benchmark. A combined analysis then reports common record-level summaries while explicitly preserving the difference between those secondary calculations and benchmark-native outcomes. The study reports benchmark performance within the stated evaluation scope. Our contributions are:
\begin{itemize}
  \item an evaluation of supplied detector scores over 1,298,747 scored records from seven public benchmark families; this is not an independent-image total because 88,000 AIGCDetectBenchmark--GenImage filename-and-label overlaps were confirmed;
  \item benchmark-native reporting together with pooled accuracy, class-rate, ranking, and error analyses, including detailed subgroup diagnostics across all seven families;
  \item dedicated benchmark sections containing official metrics, local error evidence, and protocol-aware comparisons with published detector, research-paper, and company results; and
  \item transparent coverage accounting for image-only OpenFake filtering, the AIGCDetectBenchmark decode failure, and unavailable or undecodable WildFake records, followed by a combined seven-benchmark synthesis.
\end{itemize}
\section{Related Work}

Synthetic-image detection research spans transfer-oriented detectors, benchmark design, and evaluation methodology. Reported performance depends on the detector, evaluation population, input pipeline, and aggregation rule. We review generalization, benchmark interpretation, and evidence-based image-only evaluation.

\subsection{Generalization in synthetic-image detection}

Early studies examined artifacts shared by convolutional generators. Wang et al.\ found that a classifier trained on ProGAN could transfer to several unseen CNN generators when training and preprocessing were chosen carefully \cite{wang2020cnn}. This motivated features less tied to one generator. Ojha et al.\ used representations from a pretrained vision--language model and showed that simple classification in that feature space improved transfer to unseen generative families \cite{ojha2023universal}. Other approaches use gradients \cite{tan2023gradients} or diffusion reconstruction error \cite{wang2023dire}.

Recent methods use data diversity, transformation-aware training, frequency cues, local texture, and self-supervised representations to reduce dependence on a narrow generator family \cite{yan2025sanity,liu2024fatformer,li2025safe,zhong2026sdaie}. Robustness remains conditional on generator families, real-image sources, compression, and transformations encountered during training and evaluation. A model can therefore perform strongly on a held-out split while being less reliable on naturally circulated images, unseen editing pipelines, or a different decision threshold. Generalization, robustness, calibration, and operating-point performance should be distinguished. Generalization concerns changed populations; robustness concerns input changes; calibration concerns the meaning of scores; and operating-point performance concerns the false-positive and false-negative trade-off at a selected threshold. High AP or AUROC can show score ordering without establishing that a fixed threshold is suitable for deployment.

An additional challenge is that detector evidence can arise from multiple, unstable sources. Generator-specific artifacts may be weakened by resizing, recompression, cropping, editing, or later synthesis advances. Conversely, correlations in real-image collection practices can make a detector appear effective without demonstrating a general content-level capability. Cross-generator evaluation is therefore informative only when the real and generated populations, transformations, and evaluation procedure are clearly stated. The prior literature establishes promising routes to transfer, but it also motivates reporting where that transfer is observed, where it weakens, and which metric captures the relevant behavior.

\subsection{Benchmarks and performance interpretation}

Benchmark design determines what a reported metric means. Accuracy at a fixed threshold reflects a particular operating point, whereas AP and AUROC characterize score ranking across thresholds. A record-weighted result can be dominated by large subsets, while an unweighted generator mean gives every subgroup equal influence. These choices answer different questions and should not be treated as interchangeable. Recent cross-dataset analysis similarly reports material dataset dependence in out-of-the-box AI-image detection \cite{ren2026outofbox}.

Reporting uncertainty and coverage improves interpretability. Confidence intervals describe variation; coverage accounting identifies records excluded before scoring and incomplete availability.

The benchmark families in this study provide complementary evidence. Chameleon is a high-resolution, human-curated challenge set \cite{yan2025sanity,aide_repository}; Community Forensics studies generator diversity and transfer \cite{park2025community}; and AIGIBench spans conventional generation, face manipulation, personalized generation, community imagery, and social-media imagery \cite{li2025aigibench,aigibench_repository}. MIRROR and HEDGE evaluate broader benchmark collections, but their reported Chameleon conditions differ from this study \cite{liu2026mirror,mirror_repository,wu2026hedge}. Published results are therefore contextual unless the population, preprocessing, aggregation, and training protocol align.

OpenFake version 2 separates a held-out core test from a Reddit in-the-wild test and includes image and video records; this study evaluates still images only \cite{openfake_dataset}. Community Forensics CompEval reports an unweighted generator mean across 21 generator subsets \cite{community_eval}, while AIGIBench retains separate real accuracy, fake accuracy, accuracy, and AP across 25 local subsets \cite{aigibench_repository}. AIGCDetectBenchmark spans 17 GAN and diffusion generator families under nominal, JPEG, blur, and resize conditions \cite{zhong2024patchcraft,aigcdetect_evaluation}. WildFake supplies a hierarchical generator-and-source index, although its local public snapshot is incomplete \cite{hong2025wildfake,wildfake_repository}. GenImage provides eight generator families and cross-generator and degraded-image tasks; this study evaluates its held-out \texttt{val} folders rather than reproducing the train-one-generator/test-eight protocol used by GenImage-trained methods \cite{zhu2023genimage,genimage_repository}.

These protocol differences also affect the interpretation of external baselines. Some published methods are trained on a benchmark-associated generator family, some are evaluated after image-format alignment or transformation-specific processing, and others report values from a different release or a partially overlapping test population. Such evidence can help position a result, but it does not automatically constitute a fair ranking. A careful benchmark study should distinguish matched comparisons from qualified or contextual comparisons, retain the original source conditions in the accompanying table, and avoid collapsing unlike experimental settings into a single leaderboard.

Together, these releases differ in generator coverage, real-image provenance, transformations, native aggregation, and public-release completeness. The Results section consequently retains every benchmark's native metric as its headline result, reports denominator and coverage conditions, and treats pooled measures as secondary analyses. This approach makes difficult subgroups and class asymmetries visible without claiming that one common score represents all forms of AI-generated imagery.

\subsection{Image-only evaluation}

Image-only detection matters when an image is the only available artifact. Provenance metadata, content credentials, watermarking, and platform records are evidence, but may be absent, removed through reposting or editing, unavailable, or inapplicable to legacy content. An image-only detector asks whether supplied-image evidence is consistent with the generated class under a declared decision rule. It is part of authenticity review, not a replacement for provenance or human review.

This matters for deployment. A generated-image score may support triage or review, but cannot establish authorship, intent, liability, or an image's provenance. Its usefulness depends on the decision context, consequences of false positives and false negatives, and availability of complementary evidence. Reporting image-only results separately makes those boundaries explicit while leaving broader authenticity judgments to a documented, multi-source review process.

Comparisons with published detectors, research systems, or commercial products require aligned evidence. They are most informative when benchmark release, test population, aggregation, input processing, thresholding, and training exposure are known and sufficiently similar. Where these conditions differ, comparisons remain contextual rather than controlled rankings. The paper therefore combines transparent benchmark calculations with protocol-aware comparison evidence, avoiding unsupported conclusions about universal superiority, causal mechanisms, or future real-world reliability.
\section{Methodology and Evaluation Design}

This section defines the evaluation inputs and post-scoring analysis.

\subsection{System Scope and Reproducibility}

Neural Defend describes ARCAS 1B as a multimodal foundation model for deepfake and AI-generated-content detection. This study evaluates only the supplied image-detection scores; the system description is an author attestation and was not independently verified from the supplied package.

For an image $x_i$, the evaluation records a generated-image score $s_i=f_\theta(x_i)$ and a binary decision. Neural Defend attests that one frozen detector and decision policy were used, with no benchmark training, fine-tuning, or benchmark-specific update. The supplied package does not independently establish these training, modality, or overlap claims. The analysis uses the recorded labels and scores for post-processing and statistical calculations.

The unit of analysis is a scored image record. The supplied outputs contain no provenance, text, context, or other signal fields; they do not independently establish which signals were used by the detector.

\subsection{Evaluation Design and Statistical Methods}

\subsubsection{Study Design and Record Accounting}

This subsection defines population, coverage, and decisions across benchmark families.

\paragraph{Population and labels.}

The study includes seven selected benchmark families: Chameleon, OpenFake, Community Forensics, AIGIBench, AIGCDetectBenchmark, WildFake, and GenImage. No preregistered or outcome-blind selection record exists. This convenience sample of public evaluations is not representative. Records use $y_i=0$ for real and $y_i=1$ for AI-generated or fake. ``Indexed'' denotes an eligible record before decoding, whereas ``scored'' denotes a record with a detector score. Denominators use scored records.

OpenFake contributes still images only: 127,638 source rows were considered before video rows were excluded, leaving 94,544 scored records. AIGCDetectBenchmark contains 152,598 indexed records and one decode failure. WildFake contains 714,156 rows in its official test metadata; 53,140 referenced paths were unavailable in the local public release, 661,016 paths were indexed, and 81 failed decoding, leaving 660,935 scores. Unavailable paths are associated with the real Wukong source in the local inventory. This is not an independent audit of the upstream release.

Coverage accounting is kept separate from detector error accounting. A pre-index exclusion means that an item was outside the image-only study population before scoring, whereas an unscored indexed item represents an eligible record for which decoding or score recovery failed. Neither category is counted as correct or incorrect. This matters for WildFake, where the available-case denominator differs from metadata count. Tables report scored denominators while retaining exclusions and decode failures as coverage information.

\paragraph{Decision rule and confusion counts.}

The binary rule is
\begin{equation}
s_i=f_{\theta}(x_i),\qquad
\hat y_i=\mathbb{1}[s_i\geq\tau],\qquad \tau=0.5.
\label{eq:decision}
\end{equation}

Recomputation over every saved prediction reproduced stored confusion counts under the greater-than-or-equal tie rule. One WildFake score equals 0.5 and is consequently classified as generated; no other benchmark has a threshold tie. This establishes reproducibility of saved decisions, not threshold selection. No evidence was supplied that 0.5 was chosen before benchmark access, so threshold-dependent results are interface-specific retrospective measurements.

With generated imagery as the positive class,
\begin{align}
N&=TP+TN+FP+FN,\\
\mathrm{Acc}&=\frac{TP+TN}{N},\\
\mathrm{TPR}&=\frac{TP}{TP+FN},\quad
\mathrm{TNR}=\frac{TN}{TN+FP},\\
\mathrm{Precision}&=\frac{TP}{TP+FP},\\
F_1&=\frac{2TP}{2TP+FP+FN},\\
\mathrm{BA}&=\frac{\mathrm{TPR}+\mathrm{TNR}}{2}.
\end{align}

We additionally report $\mathrm{FPR}=1-\mathrm{TNR}$ and $\mathrm{FNR}=1-\mathrm{TPR}$. Where presentation-attack terminology is useful, generated content is the attack class, so $\mathrm{APCER}=\mathrm{FNR}$, $\mathrm{BPCER}=\mathrm{FPR}$, and $\mathrm{ACER}=(\mathrm{APCER}+\mathrm{BPCER})/2$. They do not imply certification.

\subsubsection{Metrics and Aggregation}

This subsection separates fixed-threshold classification, threshold-free ranking, and benchmark-native aggregation measures.

\paragraph{Ranking metrics and native aggregation.}

Average precision is computed by grouping score ties at thresholds and summing precision weighted by recall increments,
\begin{equation}
\mathrm{AP}=\sum_k(R_k-R_{k-1})P_k.
\end{equation}

Trapezoidal area under the precision--recall curve is computed separately. This distinction matters for OpenFake: the stored field named \texttt{pr\_auc} is numerically identical to AP, but that equality does not establish historical implementation. Our supplementary recalculation gives, respectively, AP/trapezoidal PR-AUC of 99.93\%/99.93\% for \texttt{core/test}, 98.61\%/98.61\% for \texttt{reddit/test}, and 99.87\%/99.87\% for the pooled still-image records.

AUROC gives half credit to tied positive--negative score pairs:
\begin{equation}
\mathrm{AUROC}=\frac{1}{N_+N_-}\sum_{i:y_i=1}\sum_{j:y_j=0}
\left[\mathbb{1}(s_i>s_j)+\tfrac12\mathbb{1}(s_i=s_j)\right].
\end{equation}

Two-class AP and AUROC are undefined within a single-class subgroup and are shown as unavailable, never as zero.

For $S$ benchmark subsets,
\begin{equation}
\mathrm{mAcc}=\frac{1}{S}\sum_{s=1}^{S}\mathrm{Acc}_s,\qquad
\mathrm{mAP}=\frac{1}{S}\sum_{s=1}^{S}\mathrm{AP}_s.
\end{equation}

These macro quantities remain separate from pooled values. Community Forensics, AIGIBench, AIGCDetectBenchmark, and GenImage use unweighted native means. OpenFake retains separate splits.

These measures complement one another. Confusion-based quantities describe fixed-threshold behavior, whereas AP and AUROC summarize score ordering. Macro averages summarize subsets; pooled values summarize records. None establishes calibration, deployment utility, or behavior under different prevalence. Together they preserve benchmark-native outcomes and expose subgroup asymmetry, threshold sensitivity, and population weighting.

\paragraph{Uncertainty and diagnostic operating points.}

For each accuracy $\hat p=k/N$, we report a conditional 95\% Wilson interval,
\begin{align}
\mathrm{CI}_{\mathrm{Wilson}}
&=\frac{\hat p+z^2/(2N)}{1+z^2/N}\nonumber\\
&\quad{}\pm
\frac{z\sqrt{\hat p(1-\hat p)/N+z^2/(4N^2)}}{1+z^2/N},
\quad z=1.959964.
\end{align}

Intervals assume independent Bernoulli records within each evaluated benchmark population. Shared or related images and cross-benchmark overlap can violate this assumption, making them descriptive approximations. No common cluster identifier supported a cluster-aware alternative.

Supplementary equal-error rates use linear interpolation on the ROC curve. TPR at fixed FPR uses the largest achievable empirical TPR at or below that FPR. These values are retrospective test-set diagnostics, not independently validated deployment thresholds. No paired test is reported because aligned comparator predictions are unavailable.

\paragraph{Integrity checks.}

Post-processing loaded seven saved prediction CSV files; checked labels, score ranges, probability complements, and duplicate identifiers; and recomputed confusion counts, class rates, $F_1$, AP, AUROC, Wilson intervals, and subgroup aggregates. All 1,298,747 scored rows passed structural checks; duplicate identifiers and invalid rows were zero. This is a scored-record total, not an independent-image total: 88,000 AIGCDetectBenchmark--GenImage filename-and-label overlaps were confirmed. Confusion counts, primary pooled metrics, AP, AUROC, and native aggregates matched the corresponding benchmark metric files and the seven authorized master-summary entries. EER and fixed-FPR TPR were freshly computed using the stated conventions and can differ slightly from stored implementations. Model inference, training, fine-tuning, and benchmark execution were not rerun.
\section{Results}

The seven benchmark families are ordered below by decreasing pooled accuracy, using the common accuracy column in Table~\ref{tab:main_metrics}: GenImage, WildFake, AIGCDetectBenchmark, AIGIBench, Community Forensics, OpenFake, and Chameleon. Their populations and native aggregations differ, so this order is a reading aid rather than a cross-benchmark leaderboard. Each subsection presents its benchmark-specific comparison chart first and its confusion matrix second.

\subsection{GenImage Evaluation}\label{sec:genimage}
GenImage is a NeurIPS 2023 million-scale benchmark constructed from ImageNet content and advanced diffusion and GAN generators \cite{zhu2023genimage,genimage_repository}. The evaluated balanced release contains 100,000 records across ADM, BigGAN, Midjourney, VQDM, GLIDE, Stable Diffusion v1.4, Stable Diffusion v1.5, and Wukong folders. It is widely used for cross-generator testing, but many published methods train on one of its partitions.

\begin{figure}[!htb]\centering
\includegraphics[width=\columnwidth,trim=20bp 7bp 20bp 3bp,clip]{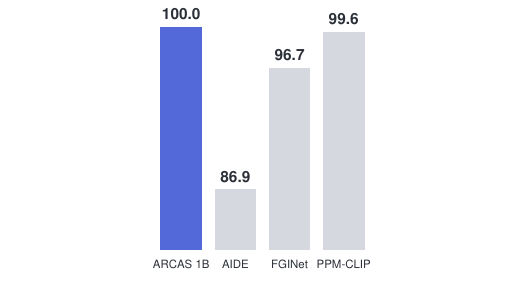}
\caption{Neural Defend ARCAS 1B and selected published GenImage results. ARCAS is evaluated on 100,000 local \texttt{val} records; external methods may use GenImage training partitions or different preprocessing. Values are contextual, not a protocol-matched ranking. Sources: AIDE, FGINet, and PPM-CLIP papers \cite{aide_repository,zhou2026fginet,wang2026ppmclip}.}\label{fig:sota_genimage}
\end{figure}

Figure~\ref{fig:sota_genimage} compares the 99.99\% eight-folder mean accuracy of ARCAS 1B with AIDE, FGINet, and PPM-CLIP. The external methods employ GenImage training, distinct crops, or other protocol choices, so the visual is contextual. It nevertheless places the evaluated result at the top of the reported score range while retaining that protocol caveat.

\begin{figure}[!htb]\centering
\includegraphics[width=\columnwidth]{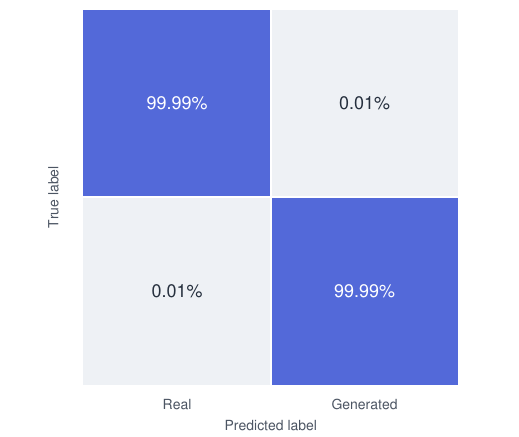}
\caption{Neural Defend ARCAS 1B confusion matrix over 100,000 scored GenImage records from the released evaluation folders \cite{zhu2023genimage,genimage_repository}.}\label{fig:confusion_genimage}
\end{figure}

Figure~\ref{fig:confusion_genimage} contains only 11 errors across 100,000 records. Midjourney contributes five, while ADM and Wukong each contribute two real-image false positives. The small count supports a high aggregate score but does not prove zero population error; the matrix keeps residual, generator-dependent errors visible.

Taken together, the chart and confusion matrix indicate consistently high performance across eight folders, not driven by one easy generator. The pattern also supports interpreting the aggregate score alongside generator-specific evidence rather than in isolation. However, the small Midjourney concentration matters because it identifies a concrete family where errors remain observable. The result should thus be read as a strong measured outcome on the released folders, with subgroup analysis retained rather than hidden by rounding to 100\%.

\subsection{WildFake Evaluation}\label{sec:wildfake}
WildFake, published at AAAI 2025, is a hierarchical benchmark spanning generators, image categories, and real-image sources \cite{hong2025wildfake,wildfake_repository}. Its hierarchy includes diffusion, GAN, and other generators such as DALL-E, Midjourney, Stable Diffusion, and StyleGAN. The available-case evaluation contains 660,935 records after path-availability and decoding checks. This is not the complete official population, so external comparisons remain contextual.

\begin{figure}[!htb]\centering
\includegraphics[width=\columnwidth,trim=20bp 7bp 20bp 3bp,clip]{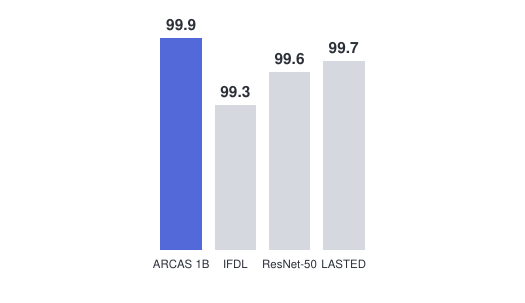}
\caption{Neural Defend ARCAS 1B and selected WildFake baselines. ARCAS is evaluated on 660,935 available, successfully decoded records; 53,140 official-index records were unavailable and 81 indexed records failed decoding. Published baselines use the complete official test and WildFake training. Values are contextual. Source: WildFake paper \cite{hong2025wildfake}.}\label{fig:sota_wildfake}
\end{figure}

Figure~\ref{fig:sota_wildfake} compares ARCAS 1B's 99.92\% available-case accuracy with published LASTED, ResNet-50, and IFDL results. The bars are close because all four scores are high, but populations differ: published rows use the complete official test and benchmark-related training, whereas ARCAS 1B uses locally available records. The comparison conveys scale, not an unconditional ranking.

\begin{figure}[!htb]\centering
\includegraphics[width=\columnwidth]{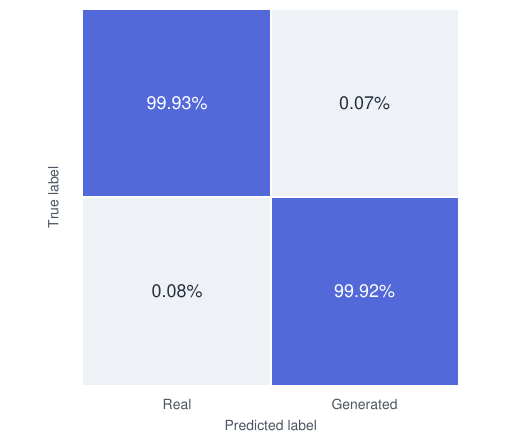}
\caption{Neural Defend ARCAS 1B available-case confusion matrix over 660,935 WildFake records \cite{hong2025wildfake,wildfake_repository}.}\label{fig:confusion_wildfake}
\end{figure}

Figure~\ref{fig:confusion_wildfake} shows a highly balanced available-case result: generated recall is 99.92\% and real-image specificity is 99.93\%. The 542 residual errors are small relative to the denominator but remain material: false negatives occur mainly among diffusion-source records, while false positives concentrate in selected real-image sources. Missing paths are a coverage limitation, not evidence of performance on unavailable records.

The two WildFake figures must be interpreted with the availability accounting in mind. The observed accuracy describes the 660,935 records that were accessible and successfully decoded, rather than all metadata rows in the official index. The near-equal recall and specificity indicate balanced behavior on that available population, while the diffusion-source false negatives identify where further diagnostic work is most useful.

\subsection{AIGCDetectBenchmark Evaluation}\label{sec:aigcdetectbenchmark}

AIGCDetectBenchmark, released with PatchCraft in 2024, covers 17 GAN and diffusion generator families \cite{zhong2024patchcraft,aigcdetect_evaluation}. The evaluation uses nominal R17. One decode failure leaves 152,597 scores. Published reports include PatchCraft, CNNSpot, NPR, AIDE, SynerDetect, SDAIE, and $S^2F$-Net on related R17 settings.

\begin{figure}[!htb]\centering
\includegraphics[width=\columnwidth,trim=20bp 7bp 20bp 3bp,clip]{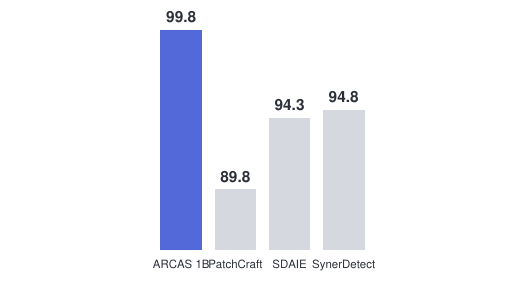}
\caption{Neural Defend ARCAS 1B and selected published results on nominal R17 AIGCDetectBenchmark mean accuracy. All values use the 17-generator mean-accuracy family; however, training and implementation conditions differ. Sources: PatchCraft, SDAIE, and SynerDetect \cite{zhong2024patchcraft,zhong2026sdaie,li2026synerdetect}.}
\label{fig:sota_aigcdetectbenchmark}
\end{figure}

Figure~\ref{fig:sota_aigcdetectbenchmark} compares ARCAS 1B's 99.79\% native mean accuracy with PatchCraft, SDAIE, and SynerDetect. The strongest plotted baseline is SynerDetect at 94.76\%. These rows share the nominal R17 metric family, although training and implementation conditions affect interpretation.

\begin{figure}[!htb]\centering
\includegraphics[width=\columnwidth]{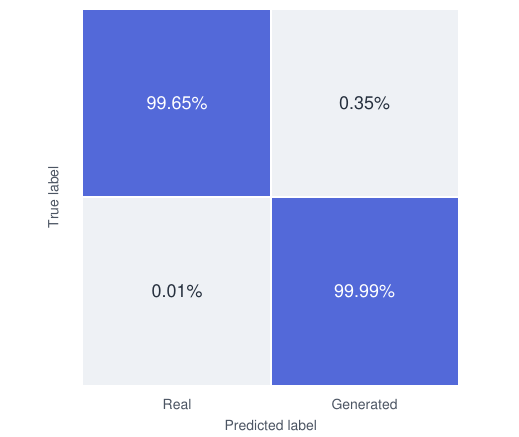}
\caption{Neural Defend ARCAS 1B confusion matrix over 152,597 scored nominal-R17 AIGCDetectBenchmark records \cite{zhong2024patchcraft,aigcdetect_evaluation}.}\label{fig:confusion_aigcdetectbenchmark}
\end{figure}

Figure~\ref{fig:confusion_aigcdetectbenchmark} shows 99.82\% pooled accuracy and near-perfect ranking performance. Residual errors are dominated by false positives: 270 real images are scored as generated, compared with seven generated images scored as real. The fixed-threshold loss is a specificity issue, not reduced generated-image sensitivity.

The comparison and confusion results answer complementary questions. The former measures equal-weighted coverage across the 17 generator families, whereas the latter shows the record-weighted operating errors at the fixed threshold. The low false-negative count supports high generated-image sensitivity; the larger false-positive count means that further calibration analysis should focus on the real-image decision boundary rather than on ranking quality.

\subsection{AIGIBench Evaluation}\label{sec:aigibench}
AIGIBench was introduced in 2025 as a robustness and generalization benchmark spanning generation, manipulation, social-media, and AI-art settings \cite{li2025aigibench,aigibench_repository}. The local release includes 25 native-resolution subsets and 212,802 scored records, covering GANs, diffusion generators, face swapping, and community imagery. Its published reference methods include retrained DGS-Net, DDA, SAFE, AIDE, Effort, and LaDeDa.

\begin{figure}[!htb]\centering
\includegraphics[width=\columnwidth,trim=20bp 7bp 20bp 3bp,clip]{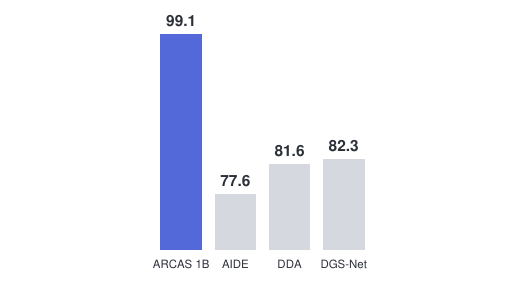}
\caption{Neural Defend ARCAS 1B and selected published AIGIBench results. ARCAS is evaluated on the supplied local snapshot; its FaceSwap and InSwap counts differ from the corrected official release. Values are contextual and not a direct leaderboard ranking. Source: AIGIBench official repository \cite{aigibench_repository}.}\label{fig:sota_aigibench}
\end{figure}

Figure~\ref{fig:sota_aigibench} places ARCAS 1B's 99.10\% native mean accuracy beside reported AIDE, DDA, and DGS-Net values. The large separation is informative but contextual: release corrections and differing FaceSwap/InSwap counts mean these rows are not strict head-to-head comparisons. The figure shows strong performance on the supplied snapshot, not an assertion that every training and population condition is identical.

\begin{figure}[!htb]\centering
\includegraphics[width=\columnwidth]{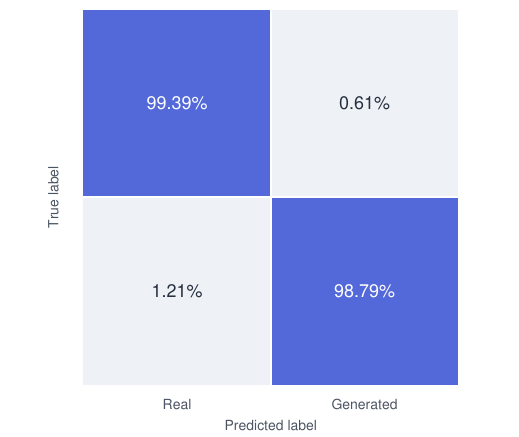}
\caption{Neural Defend ARCAS 1B confusion matrix over 212,802 AIGIBench records in the evaluated release \cite{li2025aigibench,aigibench_repository}.}\label{fig:confusion_aigibench}
\end{figure}

Figure~\ref{fig:confusion_aigibench} shows the aggregate error pattern behind the 99.10\% mean. FaceSwap and SocialRF account for a disproportionate share of errors: FaceSwap has weaker generated-image recall, while SocialRF has relatively more real-image false positives. The matrix shows that the high aggregate result is broad but not uniform across generation and manipulation modes.

The AIGIBench analysis reveals why the high mean should not be interpreted as uniform difficulty across all modes. FaceSwap is comparatively difficult because generated images are more often missed, whereas SocialRF shows a different real-image error pattern. These two distinct failure modes motivate retaining both the native mean and the confusion matrix instead of reporting one aggregate accuracy alone.

\subsection{Community Forensics Evaluation}\label{sec:community_forensics}
Community Forensics was released at CVPR 2025 to study generalization across a broad generator population \cite{park2025community,community_eval}. Its CompEval set contains 51,836 native-resolution records across 21 generator subsets and multiple real-image sources. Official reporting uses unweighted generator-level mean accuracy and mean AP, preventing the largest subsets from dominating the benchmark outcome.

\begin{figure}[!htb]\centering
\includegraphics[width=\columnwidth,trim=20bp 7bp 20bp 3bp,clip]{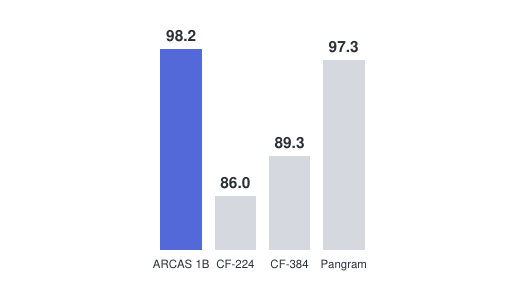}
\caption{Neural Defend ARCAS 1B and selected Community Forensics CompEval results using native macro accuracy. Pangram Image is vendor-reported, with denominator and training-overlap details not publicly disclosed. Values are contextual. Sources: Community Forensics evaluation release and Pangram Image research preview \cite{community_eval,pangram2026image}.}\label{fig:sota_communityforensics}
\end{figure}

Figure~\ref{fig:sota_communityforensics} compares the 98.17\% native mean accuracy of ARCAS 1B with the released CF-224 and CF-384 baselines and Pangram's published preview. ARCAS 1B is 0.88 points above Pangram and 8.87 points above CF-384. Pangram's denominator and training-overlap audit are undisclosed, so its row provides context rather than a controlled ranking.

\begin{figure}[!htb]\centering
\includegraphics[width=\columnwidth]{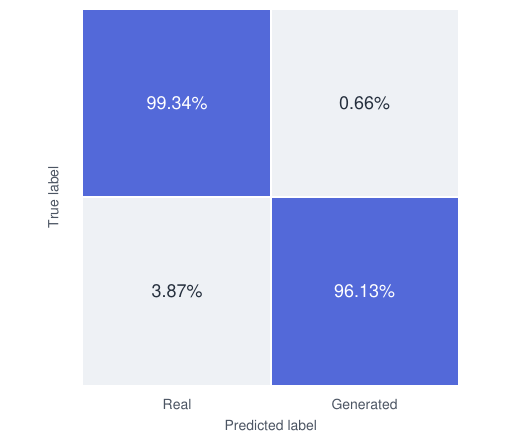}
\caption{Neural Defend ARCAS 1B confusion matrix over 51,836 Community Forensics records in CompEval \cite{park2025community,community_eval}.}\label{fig:confusion_communityforensics}
\end{figure}

Figure~\ref{fig:confusion_communityforensics} complements the generator-mean comparison with pooled record-level errors. ARCAS 1B achieves 99.92\% native mean AP, but pooled accuracy is lower because difficult large subsets receive greater weight. The matrix shows that residual errors are mostly generated images classified as real; this explains why strong threshold-free ranking can coexist with misses at the fixed 0.5 threshold.

This benchmark illustrates the difference between native and pooled aggregation. The unweighted generator mean treats each generator subset equally, while the confusion matrix reflects every scored record and therefore gives larger subsets greater influence. Reading both figures together shows that the result is strong across the generator population, but that thresholded misses remain concentrated in selected difficult subsets.

\subsection{OpenFake Evaluation}\label{sec:openfake}
OpenFake is a politically focused deepfake benchmark from 2025, containing material from proprietary and open-source generators \cite{livernoche2025openfake,openfake_dataset}. Version 2 separates a core split from naturally circulated Reddit content. After video rows are excluded, the evaluated still-image population contains 94,544 records. The original paper's SwinV2/OpenFake result is a version-1 baseline, so it is contextual rather than a matched version-2 ranking.

\begin{figure}[!htb]\centering
\includegraphics[width=\columnwidth,trim=20bp 7bp 20bp 3bp,clip]{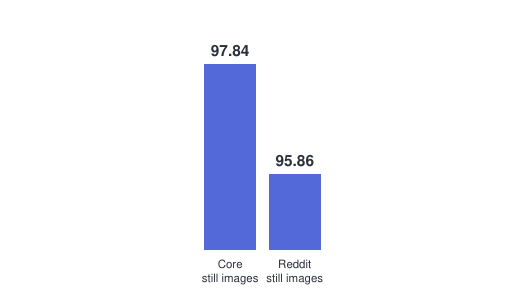}
\caption{Neural Defend ARCAS 1B accuracy on OpenFake version-2 still-image evaluation splits \cite{livernoche2025openfake,openfake_dataset}.}\label{fig:openfake_split}
\end{figure}

OpenFake has no directly matched version-2 SOTA baseline, so Figure~\ref{fig:openfake_split} compares the two official still-image evaluation splits. Accuracy declines from 97.84\% on the held-out core split to 95.86\% on Reddit. The chart preserves this transfer difference rather than masking it in a pooled score.

\begin{figure}[!htb]\centering
\includegraphics[width=\columnwidth]{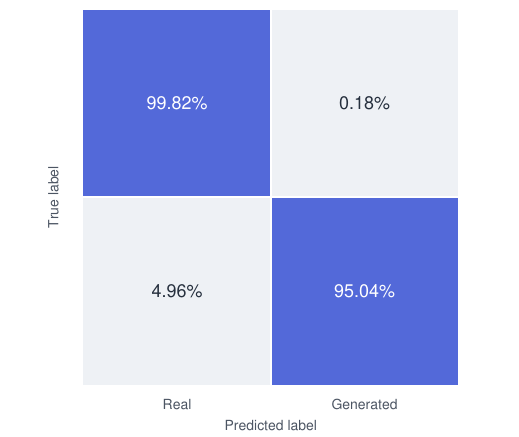}
\caption{Neural Defend ARCAS 1B confusion matrix pooled across the OpenFake still-image core and Reddit splits \cite{livernoche2025openfake,openfake_dataset}.}\label{fig:confusion_openfake}
\end{figure}

Figure~\ref{fig:confusion_openfake} summarizes 97.65\% pooled accuracy, 99.88\% AUROC, and 97.35\% $F_1$. Generated-image recall decreases on naturally circulated material, whereas real-image specificity remains high. The main transfer cost is therefore missed generated images rather than an increase in real-image false alarms.

The OpenFake results are best understood as a transfer analysis. The split chart isolates the decline on naturally circulated Reddit imagery, while the confusion matrix shows that the pooled result remains driven by missed generated images rather than a broad false-positive problem. Because no matched version-2 published detector score is available, this subsection avoids presenting the outcome as a SOTA ranking.

\subsection{Chameleon Evaluation}\label{sec:chameleon}
Chameleon, introduced with AIDE at ICLR 2025, is a human-curated, high-resolution benchmark designed to test detector generalization \cite{yan2025sanity,aide_repository}. The supplied test set contains 26,033 records: 14,863 real and 11,170 generated images. It uses ordinary accuracy as its native headline measure. The published landscape includes AIDE, SimLBR, FGINet, TAP, MIRROR, and HEDGE, although training data and preprocessing conditions differ.

\begin{figure}[!htb]\centering
\includegraphics[width=\columnwidth,trim=20bp 7bp 20bp 3bp,clip]{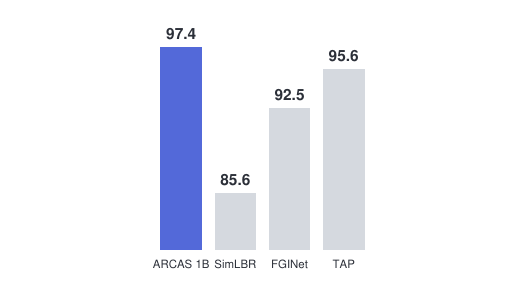}
\caption{Neural Defend ARCAS 1B and selected published Chameleon accuracy results. ARCAS, SimLBR, FGINet, and TAP use differing training data, foundation-model initialization, and/or preprocessing. Values are contextual, not a controlled ranking. Sources: AIDE, SimLBR, FGINet, and TAP papers \cite{aide_repository,dhakal2026simlbr,zhou2026fginet,abdullah2026tap}.}\label{fig:sota_chameleon}
\end{figure}

Figure~\ref{fig:sota_chameleon} compares the reported Chameleon accuracy of ARCAS 1B with selected published values. ARCAS 1B reaches 97.44\%, exceeding the plotted TAP (95.64\%), FGINet (92.50\%), and SimLBR (85.57\%) rows. This is descriptive rather than a claim of identical protocols: several external methods use benchmark-related training material or different preprocessing pipelines.

\begin{figure}[!htb]\centering
\includegraphics[width=\columnwidth]{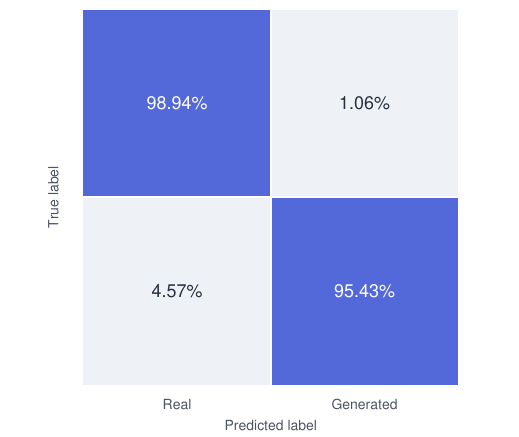}
\caption{Neural Defend ARCAS 1B confusion matrix over 26,033 Chameleon records from the released benchmark \cite{yan2025sanity,aide_repository}.}\label{fig:confusion_chameleon}
\end{figure}

Figure~\ref{fig:confusion_chameleon} shows that the 97.44\% result is not perfectly symmetric across classes. Generated-image recall is 95.43\%, while real-image specificity is 98.94\%; consequently, false negatives are more common than false positives. The matrix identifies generated-image sensitivity, rather than real-image rejection, as the principal residual error source.

The Chameleon result is the lowest pooled accuracy in this paper's selected set, so its error profile is particularly informative. The difference between recall and specificity indicates that difficult generated images are the limiting class under the shared threshold. The model-comparison figure identifies a high observed score relative to the plotted literature, while the matrix prevents that comparison from obscuring the remaining generated-image misses.

\FloatBarrier
\subsection{Cross-benchmark analysis}\label{sec:cross_benchmark}
Table~\ref{tab:main_metrics} follows the same decreasing pooled-accuracy sequence as the result subsections. It consolidates accuracy, AUROC, $F_1$, precision, recall, specificity, APCER, and BPCER for all seven benchmark populations. It provides a shared operating view, but is not a single leaderboard: Community Forensics, AIGIBench, and AIGCDetectBenchmark use native generator-mean measures, whereas WildFake is explicitly available-case.
\begin{table*}[t]
\centering
\caption{Neural Defend ARCAS 1B per-dataset performance metrics. APCER is the generated-image error rate ($1-\mathrm{Recall}$) and BPCER is the real-image error rate ($1-\mathrm{Specificity}$). EER/ACER is not reported because no benchmark-specific operating threshold is selected for that measure.}
\label{tab:main_metrics}
\scriptsize
\setlength{\tabcolsep}{3pt}
\begin{tabular}{lrrrrrrrr}
\toprule
Dataset & Accuracy & AUROC & $F_1$ & Precision & Recall & Specificity & APCER & BPCER \\
\midrule
GenImage & 99.99 & 100.00 & 99.99 & 99.99 & 99.99 & 99.99 & 0.01 & 0.01 \\
WildFake & 99.92 & 100.00 & 99.95 & 99.98 & 99.92 & 99.93 & 0.08 & 0.07 \\
AIGCDetectBenchmark & 99.82 & 100.00 & 99.82 & 99.65 & 99.99 & 99.65 & 0.01 & 0.35 \\
AIGIBench & 99.09 & 99.91 & 99.09 & 99.38 & 98.79 & 99.39 & 1.21 & 0.61 \\
Community Forensics & 97.74 & 99.72 & 97.70 & 99.32 & 96.13 & 99.34 & 3.87 & 0.66 \\
OpenFake & 97.65 & 99.88 & 97.35 & 99.77 & 95.04 & 99.82 & 4.96 & 0.18 \\
Chameleon & 97.44 & 99.51 & 96.97 & 98.55 & 95.43 & 98.94 & 4.57 & 1.06 \\
\bottomrule
\end{tabular}
\end{table*}

Table~\ref{tab:main_metrics} makes ordering fully reproducible: its first column is sorted from 99.99\% to 97.44\% accuracy, exactly matching the sequence of the subsections above. The threshold-free AUROC column remains near ceiling across all rows, whereas the recall, specificity, APCER, and BPCER columns expose distinct fixed-threshold behavior. In particular, OpenFake and Chameleon have the largest APCER values, while AIGCDetectBenchmark has a comparatively larger BPCER.

Across the table rows, AUROC is high and APCER/BPCER are low, but the residual-error type changes by benchmark. Chameleon, OpenFake, and Community Forensics have lower generated-image recall than real-image specificity, so missed generated images dominate their remaining errors. AIGCDetectBenchmark reverses that pattern: false positives account for most of its small residual. Similar accuracy scores can therefore conceal materially different decision risks.

The populations are also unequal. WildFake supplies the largest available-case denominator, so a micro-average would be heavily influenced by it; equal weighting of benchmark rows instead gives small and large datasets the same influence. Cross-benchmark claims are limited to observed public evaluation records and do not establish future robustness, attribution, or deployment performance.

\begin{figure*}[t]\centering
\includegraphics[width=0.94\textwidth]{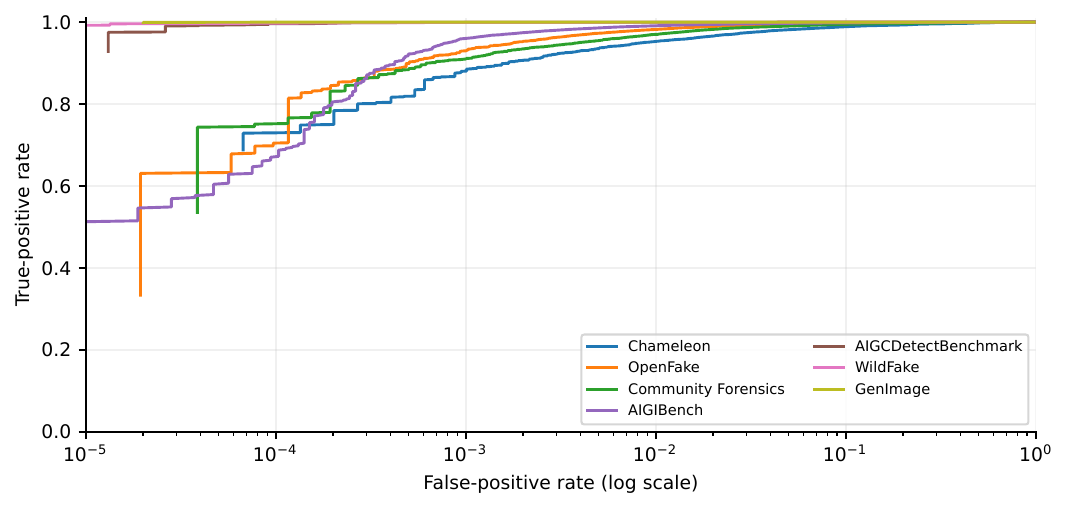}
\caption{Neural Defend ARCAS 1B low-false-positive-rate ROC curves across evaluated benchmark populations. This combined view compares discrimination where false positives are constrained; benchmark sources are cited in the corresponding subsections.}\label{fig:low_fpr_roc}
\end{figure*}

Figure~\ref{fig:low_fpr_roc} compares low-false-positive-rate ROC behavior across the benchmark populations. It is most useful where false positives carry a high review cost: curves nearer the upper-left region retain generated-image sensitivity while constraining real-image errors. The plot supplements the table's fixed 0.5 operating point rather than selecting a new threshold for any benchmark.

\section{Discussion}

\subsection{Interpretation of the evidence}

The evaluation supports strong measured performance for the supplied detector scores on the selected records, while the subgroup findings qualify the aggregate picture. Neural Defend's statements that the detector is frozen, multimodal or foundational, image-only, and free of benchmark-training overlap are author attestations, not independently established by the supplied package. Firefly Image 3, Hourglass, FaceSwap, SocialRF, and OpenFake Reddit expose concentrated errors that broad averages conceal. Threshold-free ranking and fixed-threshold classification also diverge in some groups: high AP does not guarantee equally high generated recall at $\tau=0.5$. The study measures behavior; it does not identify its cause, so claims about model scale, training data, components, or optimization would be unsupported.

\subsection{Operational interpretation}

In trust-and-safety and media-forensics workflows, a high TNR reduces unnecessary escalation of genuine content, while high TPR reduces generated content missed at the selected operating point. Their costs are application-specific. Chameleon's lower TPR and AIGCDetectBenchmark's larger false-positive than false-negative count show why an operator should not use aggregate accuracy alone to set policy. The saved scores may support later threshold studies, but a threshold selected retrospectively on these test sets would require validation on a separate, operationally representative population.

The detector is best treated as one signal in a layered content-authenticity process. Scores can prioritize expert review, complement provenance metadata, or prompt additional forensic checks. They should not by themselves establish authorship, intent to deceive, legal liability, or grounds for punitive action. A binary generated/real score also does not attribute an image to a particular generator.

Operational deployment also requires monitoring after release. Changes in image sources, compression practices, or generator capabilities can alter error rates even when benchmark accuracy remains high. Teams should therefore log decision context, review disagreement cases, and reassess thresholds against current representative samples. Such monitoring supports accountable use without converting a benchmark score into a standalone authenticity judgment.

\subsection{Meaning of the comparison evidence}

The benchmark-local comparison tables are evidence maps, not a single leaderboard. They identify the conditions under which a published detector, a company result, or a benchmark baseline can be compared with ARCAS. The most aligned panels are full-split Chameleon ordinary accuracy, released-CompEval mAcc/mAP, and nominal R17 AIGCDetectBenchmark mAcc. Other rows remain contextual because of benchmark exposure, training, population, preprocessing, version, or missingness differences. This distinction is essential: numerical separation in a table does not isolate method quality under equal resources or establish an unrestricted state of the art.

\subsection{Deployment validation and score governance}

Benchmark scores require separate operational validation. Select thresholds on a population representing intended sources, transformations, prevalence, review costs, and capacity, not retrospectively on the seven test sets. Pre-specify use case and error costs, fix the threshold, then assess it on a later holdout sample.

Ranking performance does not establish calibration. Validation should report reliability curves, Brier score, and expected calibration error (ECE), with conventions declared in advance. Any recalibration uses validation data only and a later holdout.

Scores should route content, not make final authenticity judgments. High-confidence cases may trigger provenance checks, intermediate cases review, and low-confidence cases sampling-based audit. The policy should define review evidence, overrides, disagreement, and outcome retention.

Post-release monitoring should track score drift, calibration, coverage failures, and false positives/negatives by source, generator family, and transformation where labels exist. New generators, compression, reposting, and real-image sources can alter errors. These studies are not performed here.

Validation should reflect acquisition paths, compression, reposting artifacts, prevalence, labeling uncertainty. Calibration reports should declare conventions and assess subgroups. Escalation policies should record reviewers, evidence, and overrides. Monitoring should specify review intervals, drift triggers, and versioned threshold revisions. These are proposed deployment procedures, not evidence supplied here.

Validation plans should define threshold ownership, evidence retention, and conditions under which material disagreement or coverage changes pause automated routing pending review by responsible staff.

\section{Conclusion}

This study evaluates supplied Neural Defend ARCAS 1B detector scores across seven selected public benchmark families. It reports each benchmark in its native evaluation setting, supplements those outcomes with common pooled and subgroup analyses, and confines external comparisons to evidence whose metric and protocol can be interpreted responsibly. The resulting record shows strong measured performance on the supplied evaluation records, but it also preserves the error patterns, coverage gaps, protocol differences, and confirmed cross-benchmark filename-and-label overlap that a single aggregate score would hide. Accordingly, the paper offers a reproducible analysis of benchmark results rather than a claim that AI-image detection is solved or universally reliable. Future work should test independently versioned deployments on temporally distinct and operationally representative populations, with calibration and decision thresholds evaluated separately from benchmark ranking metrics.

Interpretation should remain tied to released records and declared denominators, while deployment requires prospective validation, documented governance, and monitoring for changes in image sources, generator families, and transformations over time.

\paragraph{Limitations}

The selected families do not represent all generators, transformations, contexts, or future releases. Version differences, one AIGCDetectBenchmark decode failure, and substantial WildFake path unavailability further limit interpretation. Wilson intervals are conditional on independent records despite possible shared sources and cross-benchmark overlap, including approximately 88,000 shared filename-and-label signatures reported between AIGCDetectBenchmark and GenImage.

The study does not establish calibration, temporal drift, adversarial robustness, fairness, attribution, or operational costs, and it includes no external blind evaluation. Dataset terms remain governed by their providers.

ARCAS 1B was evaluated through supplied detector scores across seven selected public benchmark families. The benchmark-native results, local comparison evidence, common pooled analysis, and subgroup diagnostics show measured performance while retaining meaningful differences in populations, error patterns, coverage, and cross-benchmark overlap. The evidence supports a carefully scoped account of benchmark performance on these evaluation records, not a claim that detection is solved or that the system is universally reliable.
\section{Declarations}

\paragraph{Data, materials, and reproducibility.}
The benchmark datasets are available from their providers under current terms. This manuscript does not redistribute benchmark images. Aggregate tables, vector figures, and LaTeX source accompany the authors' package. Per-record predictions and full-precision scores are retained and are not included in the submission archive. ARCAS 1B architecture, weights, private training data, training code, and an inference API are not released.

The private authors' package includes post-processing, numerical reconciliation, source ledger, claim-to-evidence matrix, comparison eligibility decisions, and build/visual-QA records. Keeping these records private avoids publishing per-record identifiers, input paths, or unapproved materials. The submission archive reproduces the manuscript presentation and aggregate calculations, not the proprietary detector or benchmark content. Independent users may inspect the released LaTeX source, tables, and figures; restricted identifiers and proprietary operational information remain subject to authors' and provider approval.

\paragraph{Conflict of interest and funding.}
The authors are affiliated with Neural Defend Inc, the developer of the proprietary system evaluated in this paper. The study therefore concerns a commercial system in which each author's employer has an interest; this conflict-of-interest statement applies to all four authors. No funding information was supplied for this manuscript; no funding claim is made.

\paragraph{Ethics and benchmark use.}
This work reports secondary analysis of existing public benchmark evaluation records and did not conduct a new human-participant study or collect new participant data. This statement is not an institutional review determination. Benchmark reuse remains subject to source terms and applicable institutional requirements.

\paragraph{Author contributions and AI assistance.}
A detailed CRediT allocation has not been finalized and is therefore not asserted. Generative AI tools assisted with drafting, literature lookup, analysis scripting, citation checks, and consistency review. All numerical outputs were checked against saved evaluation records and cited sources as described in the private verification package. The authors remain responsible for the manuscript, source interpretation, disclosures, and submission decisions. This statement should be adapted if a later venue specifies a different disclosure policy.

\paragraph{Literature cutoff.}
The source and comparison search was last updated on \searchcutoff. A later submission should recheck paper versions, benchmark cards, licenses, and venue status.

\clearpage
\bibliographystyle{plainnat}
\bibliography{references}

\clearpage
\onecolumn
\appendix
\raggedbottom
\raggedright
\section{Additional Results}
\label{app:additional_results}

Table~\ref{tab:sota_comparison} places ARCAS 1B beside one selected prior result for every evaluated benchmark. The prior-method values are cited within the table and retain their original metric, split, and protocol context. The comparison should therefore be read as traceable evidence by benchmark, not as a pooled cross-paper ranking.
\begin{center}
\captionof{table}{Neural Defend ARCAS 1B versus selected prior methods across the evaluated benchmark families. Values retain each source's reported metric and protocol; rows are not a unified leaderboard.}
\label{tab:sota_comparison}
\small
\setlength{\tabcolsep}{4pt}
\begin{tabular}{p{0.16\textwidth} p{0.17\textwidth} p{0.10\textwidth} p{0.24\textwidth} p{0.24\textwidth}}
\toprule
Benchmark & Neural Defend ARCAS 1B & Prior result & Prior method & Comparison context \\
\midrule
Chameleon & 97.44 Acc. & 95.64 Acc. & TAP PE-ViT-G + TAP\newline\cite{abdullah2026tap} & Full split; qualified training protocol \\
OpenFake & 97.65 Acc. & 99.20 $F_1$ & SwinV2/OpenFake\newline\cite{livernoche2025openfake} & Version-1 result; not directly matched \\
Community Forensics & 98.17 mAcc & 97.29 mAcc & Pangram Image\newline\cite{pangram2026image} & Vendor-reported CompEval result \\
AIGIBench & 99.10 mAcc & 82.30 mAcc & DGS-Net\newline\cite{aigibench_repository} & Repository Setting-II result \\
AIGCDetectBenchmark & 99.79 mAcc & 94.76 mAcc & SynerDetect\newline\cite{li2026synerdetect} & Nominal clean R17 \\
WildFake & 99.92 Acc. & 99.70 Acc. & LASTED\newline\cite{hong2025wildfake} & Complete test versus available-case evaluation \\
GenImage & 99.99 mAcc & 99.60 mAcc & PPM-CLIP\newline\cite{wang2026ppmclip} & Eight-folder mean; SD1.4 training \\
\bottomrule
\end{tabular}
\end{center}

The comparison table shows that the strongest directly comparable evidence occurs for nominal R17 AIGCDetectBenchmark and released CompEval reporting. OpenFake remains a version-mismatched comparison, while WildFake and several GenImage rows differ in available population or training exposure. These distinctions explain why the benchmark-specific Results subsections remain the primary location for interpretation.

Table~\ref{tab:generator_performance} reports selected generator-level results from releases that expose named generator labels. It includes DALL-E 3, FLUX.1-dev, Midjourney, Stable Diffusion, and Adobe Firefly variants; GPT and Gemini are not included because the evaluated benchmark releases did not provide them as named generator labels.
\begin{center}
\captionof{table}{Neural Defend ARCAS 1B selected generator-level performance. GPT and Gemini were not present as named generator labels in the evaluated releases.}
\label{tab:generator_performance}
\small
\begin{tabular}{l l r r r r}
\toprule
Generator & Benchmark & $N$ & Accuracy & AP & Recall \\
\midrule
DALL-E 3 & AIGIBench & 8,002 & 99.70 & 99.96 & 99.60 \\
FLUX.1-dev & AIGIBench & 9,000 & 99.82 & 100.00 & 99.93 \\
Midjourney & AIGIBench & 6,000 & 99.55 & 100.00 & 99.27 \\
Midjourney & GenImage & 12,000 & 99.96 & 100.00 & 99.93 \\
Stable Diffusion v1.4 & GenImage & 12,000 & 100.00 & 100.00 & 100.00 \\
Stable Diffusion v1.5 & GenImage & 16,000 & 99.99 & 100.00 & 100.00 \\
Firefly Image 2 & Community Forensics & 3,842 & 96.15 & 99.85 & 92.45 \\
Firefly Image 3 & Community Forensics & 3,896 & 90.71 & 99.38 & 81.78 \\
\bottomrule
\end{tabular}
\end{center}

The generator-level values show that high aggregate accuracy does not imply identical difficulty across every family. For example, the Firefly rows have lower recall than the Stable Diffusion rows, while Midjourney results differ by benchmark population. These per-generator records complement the main-text confusion matrices by locating aggregate errors in identifiable generator families.

\section{Complete Legacy Subset Tables}
\label{app:legacy_subset_tables}

The following tables are retained from the earlier appendix PDF and report every recoverable benchmark subset from the saved predictions. Rates and AP are percentages; intervals are conditional 95\% Wilson intervals where reported. A dash denotes a quantity that is undefined for a single-class subset, and an observed value of 100.0000\% does not imply perfect population performance.

Table~\ref{tab:Chameleon_subsets} separates the real and generated Chameleon subsets. Table~\ref{tab:CommunityForensics_subsets} gives the complete generator-level CompEval record, including the more difficult Firefly and Hourglass groups.
\begin{center}
\begin{minipage}{\textwidth}
\centering
\captionsetup{type=table,hypcap=false}
\captionof{table}{Chameleon: complete available subset statistics. Rates and AP are percentages; a dash denotes a metric undefined for a single-class subset. All counts are scored records.}
\label{tab:Chameleon_subsets}
\small
\setlength{\tabcolsep}{3.5pt}
\begin{tabular*}{\textwidth}{@{\extracolsep{\fill}}lrrrrrrr}
\toprule
Subset & $N$ & Acc. (\%) & AP & TNR & TPR & FP & FN \\
\midrule
0\_real & 14,863 & 98.9437 & -- & 98.9437 & -- & 157 & 0 \\
1\_fake & 11,170 & 95.4342 & -- & -- & 95.4342 & 0 & 510 \\
\bottomrule
\end{tabular*}
\end{minipage}
\end{center}
\medskip

\begin{center}
\begin{minipage}{\textwidth}
\centering
\captionsetup{type=table,hypcap=false}
\captionof{table}{Community Forensics: complete available subset statistics. Rates and AP are percentages; a dash denotes a metric undefined for a single-class subset. All counts are scored records.}
\label{tab:CommunityForensics_subsets}
\small
\setlength{\tabcolsep}{3.5pt}
\begin{tabular*}{\textwidth}{@{\extracolsep{\fill}}lrrrrrrr}
\toprule
Subset & $N$ & Acc. (\%) & AP & TNR & TPR & FP & FN \\
\midrule
DFGAN & 2,000 & 100.0000 & 100.0000 & 100.0000 & 100.0000 & 0 & 0 \\
Dalle2 & 2,000 & 96.9500 & 99.9090 & 99.7000 & 94.2000 & 3 & 58 \\
Dalle3 & 2,000 & 99.8000 & 99.9980 & 99.7000 & 99.9000 & 3 & 1 \\
DeciDiffusionV2 & 2,000 & 99.3000 & 100.0000 & 98.6000 & 100.0000 & 14 & 0 \\
FLUX-dev & 2,000 & 99.5000 & 99.9705 & 99.6000 & 99.4000 & 4 & 6 \\
FLUX-schnell & 2,000 & 99.6000 & 99.9892 & 99.7000 & 99.5000 & 3 & 5 \\
Firefly\_Image2 & 3,842 & 96.1478 & 99.8494 & 99.8438 & 92.4518 & 3 & 145 \\
Firefly\_Image3 & 3,896 & 90.7084 & 99.3774 & 99.6407 & 81.7762 & 7 & 355 \\
GALIP & 2,000 & 99.5000 & 99.9997 & 99.0000 & 100.0000 & 10 & 0 \\
Hourglass & 4,000 & 93.2000 & 99.5708 & 99.9500 & 86.4500 & 1 & 271 \\
IdeogramV1 & 2,000 & 97.8500 & 99.9324 & 99.6000 & 96.1000 & 4 & 39 \\
IdeogramV2 & 2,000 & 96.5500 & 99.9063 & 99.7000 & 93.4000 & 3 & 66 \\
Imagen3 & 2,114 & 98.7701 & 99.9607 & 99.7162 & 97.8240 & 3 & 23 \\
LCM\_lora\_sdv15 & 2,000 & 99.0000 & 99.9890 & 98.8000 & 99.2000 & 12 & 8 \\
LCM\_lora\_sdxl & 2,000 & 99.2500 & 99.9997 & 98.5000 & 100.0000 & 15 & 0 \\
LCM\_lora\_ssd1b & 2,000 & 99.4500 & 99.9998 & 98.9000 & 100.0000 & 11 & 0 \\
MidjourneyV5\_2 & 3,986 & 99.7240 & 99.9994 & 99.4982 & 99.9498 & 10 & 1 \\
MidjourneyV6\_1 & 3,998 & 99.3497 & 99.9781 & 99.5998 & 99.0995 & 8 & 18 \\
kandinsky\_2\_2 & 2,000 & 98.8500 & 100.0000 & 97.7000 & 100.0000 & 23 & 0 \\
kvikontent\_midjourney\_v6 & 2,000 & 99.2000 & 99.9929 & 99.0000 & 99.4000 & 10 & 6 \\
stable\_cascade & 2,000 & 98.8000 & 99.9999 & 97.6000 & 100.0000 & 24 & 0 \\
\bottomrule
\end{tabular*}
\end{minipage}
\end{center}
\medskip

Table~\ref{tab:AIGIBench_subsets} and Table~\ref{tab:AIGCDetectBenchmark_subsets} retain all available AIGIBench and AIGCDetectBenchmark generator rows. They provide the detailed basis for the aggregate scores and confusion-matrix interpretations in the Results section.
\enlargethispage{4\baselineskip}
\begin{center}
\begin{minipage}{\textwidth}
\centering
\captionsetup{type=table,hypcap=false}
\captionof{table}{AIGIBench: complete available subset statistics. Rates and AP are percentages; a dash denotes a metric undefined for a single-class subset. All counts are scored records.}
\label{tab:AIGIBench_subsets}
\small
\setlength{\tabcolsep}{3.5pt}
\begin{tabular*}{\textwidth}{@{\extracolsep{\fill}}lrrrrrrr}
\toprule
Subset & $N$ & Acc. (\%) & AP & TNR & TPR & FP & FN \\
\midrule
BLIP & 9,000 & 99.6889 & 100.0000 & 99.3778 & 100.0000 & 28 & 0 \\
BlendFace & 9,000 & 97.9111 & 99.8883 & 99.8222 & 96.0000 & 8 & 180 \\
CommunityAI & 12,000 & 98.1167 & 99.7711 & 98.9667 & 97.2667 & 62 & 164 \\
DALLE-3 & 8,002 & 99.7001 & 99.9607 & 99.8001 & 99.6000 & 8 & 16 \\
E4S & 9,000 & 99.2222 & 99.9373 & 99.7111 & 98.7333 & 13 & 57 \\
FLUX1-dev & 9,000 & 99.8222 & 99.9963 & 99.7111 & 99.9333 & 13 & 3 \\
FaceSwap & 8,900 & 95.8315 & 99.3806 & 99.5556 & 92.0227 & 20 & 351 \\
GLIDE & 9,000 & 99.8222 & 100.0000 & 99.6444 & 100.0000 & 16 & 0 \\
IP\_Adapter & 9,000 & 99.7000 & 99.9976 & 99.4667 & 99.9333 & 24 & 3 \\
Imagen3 & 9,000 & 99.5222 & 99.9845 & 99.7556 & 99.2889 & 11 & 32 \\
InSwap & 8,900 & 98.5056 & 99.8665 & 99.6222 & 97.3636 & 17 & 116 \\
Infinite\_ID & 9,000 & 99.7667 & 100.0000 & 99.5333 & 100.0000 & 21 & 0 \\
InstantID & 9,000 & 99.6444 & 99.9960 & 99.6000 & 99.6889 & 18 & 14 \\
Midjourney & 6,000 & 99.5500 & 99.9960 & 99.8333 & 99.2667 & 5 & 22 \\
PhotoMaker & 9,000 & 98.3222 & 99.9508 & 99.6222 & 97.0222 & 17 & 134 \\
ProGAN & 8,000 & 98.1500 & 99.9998 & 96.3000 & 100.0000 & 148 & 0 \\
R3GAN & 9,000 & 99.7333 & 99.9999 & 99.4667 & 100.0000 & 24 & 0 \\
SD3 & 9,000 & 99.8000 & 99.9995 & 99.6667 & 99.9333 & 15 & 3 \\
SDXL & 9,000 & 99.8333 & 100.0000 & 99.6667 & 100.0000 & 15 & 0 \\
SimSwap & 9,000 & 98.7000 & 99.9021 & 99.4667 & 97.9333 & 24 & 93 \\
SocialRF & 6,000 & 96.8667 & 98.7099 & 96.2667 & 97.4667 & 112 & 76 \\
StyleGAN-XL & 9,000 & 99.8000 & 99.9999 & 99.6000 & 100.0000 & 18 & 0 \\
StyleGAN3 & 9,000 & 99.7667 & 99.9984 & 99.9333 & 99.6000 & 3 & 18 \\
StyleSwim & 9,000 & 99.8444 & 100.0000 & 99.6889 & 100.0000 & 14 & 0 \\
WFIR & 2,000 & 100.0000 & 100.0000 & 100.0000 & 100.0000 & 0 & 0 \\
\bottomrule
\end{tabular*}
\end{minipage}
\end{center}
\medskip
\begin{center}
\begin{minipage}{\textwidth}
\centering
\captionsetup{type=table,hypcap=false}
\captionof{table}{AIGCDetectBenchmark: complete available subset statistics. Rates and AP are percentages; a dash denotes a metric undefined for a single-class subset. All counts are scored records.}
\label{tab:AIGCDetectBenchmark_subsets}
\small
\setlength{\tabcolsep}{3.5pt}
\begin{tabular*}{\textwidth}{@{\extracolsep{\fill}}lrrrrrrr}
\toprule
Subset & $N$ & Acc. (\%) & AP & TNR & TPR & FP & FN \\
\midrule
ADM & 12,000 & 99.9833 & 100.0000 & 99.9667 & 100.0000 & 2 & 0 \\
DALLE2 & 1,999 & 99.8499 & 99.9990 & 99.8999 & 99.8000 & 1 & 2 \\
Glide & 12,000 & 100.0000 & 100.0000 & 100.0000 & 100.0000 & 0 & 0 \\
Midjourney & 12,000 & 99.9583 & 100.0000 & 99.9833 & 99.9333 & 1 & 4 \\
VQDM & 12,000 & 100.0000 & 100.0000 & 100.0000 & 100.0000 & 0 & 0 \\
biggan & 4,000 & 99.4750 & 100.0000 & 98.9500 & 100.0000 & 21 & 0 \\
cyclegan & 2,642 & 99.8864 & 99.9999 & 99.8486 & 99.9243 & 2 & 1 \\
gaugan & 10,000 & 99.9200 & 100.0000 & 99.8400 & 100.0000 & 8 & 0 \\
progan & 8,000 & 98.1500 & 99.9998 & 96.3000 & 100.0000 & 148 & 0 \\
sd\_xl & 4,000 & 99.9500 & 100.0000 & 99.9000 & 100.0000 & 2 & 0 \\
stable\_diffusion\_v\_1\_4 & 12,000 & 100.0000 & 100.0000 & 100.0000 & 100.0000 & 0 & 0 \\
stable\_diffusion\_v\_1\_5 & 16,000 & 99.9938 & 100.0000 & 99.9875 & 100.0000 & 1 & 0 \\
stargan & 3,998 & 99.6498 & 100.0000 & 99.2996 & 100.0000 & 14 & 0 \\
stylegan & 11,982 & 99.8832 & 100.0000 & 99.7663 & 100.0000 & 14 & 0 \\
stylegan2 & 15,976 & 99.6620 & 99.9998 & 99.3240 & 100.0000 & 54 & 0 \\
whichfaceisreal & 2,000 & 100.0000 & 100.0000 & 100.0000 & 100.0000 & 0 & 0 \\
wukong & 12,000 & 99.9833 & 100.0000 & 99.9667 & 100.0000 & 2 & 0 \\
\bottomrule
\end{tabular*}
\end{minipage}
\end{center}
\medskip

WildFake is preserved at both broad-family and recoverable-source levels. Table~\ref{tab:WildFake_subsets} reports diffusion, GAN, other-generated, and real branches; Tables~\ref{tab:WildFake_generated_sources} and \ref{tab:WildFake_real_sources} identify the source-level false-negative and false-positive patterns.
\begin{center}
\begin{minipage}{\textwidth}
\centering
\captionsetup{type=table,hypcap=false}
\captionof{table}{WildFake: complete available subset statistics. Rates and AP are percentages; a dash denotes a metric undefined for a single-class subset. All counts are scored records.}
\label{tab:WildFake_subsets}
\small
\setlength{\tabcolsep}{3.5pt}
\begin{tabular*}{\textwidth}{@{\extracolsep{\fill}}lrrrrrrr}
\toprule
Subset & $N$ & Acc. (\%) & AP & TNR & TPR & FP & FN \\
\midrule
Diffusion\_based & 377,290 & 99.8972 & -- & -- & 99.8972 & 0 & 388 \\
GAN\_based & 98,614 & 99.9605 & -- & -- & 99.9605 & 0 & 39 \\
Other\_based & 35,478 & 99.9915 & -- & -- & 99.9915 & 0 & 3 \\
Real & 149,553 & 99.9251 & -- & 99.9251 & -- & 112 & 0 \\
\bottomrule
\end{tabular*}
\end{minipage}
\end{center}
\medskip

\begin{center}
\begin{minipage}{\textwidth}
\centering
\captionsetup{type=table,hypcap=false}
\captionof{table}{WildFake generated-source results recovered from the documented hierarchy in saved record identifiers. Each row is generated-only, so only TPR and false negatives are defined.}
\label{tab:WildFake_generated_sources}
\small
\setlength{\tabcolsep}{3.5pt}
\begin{tabular*}{\textwidth}{@{\extracolsep{\fill}}lrrrr}
\toprule
Source & $N$ & FN & TPR & 95\% CI \\
\midrule
ADM & 31,005 & 1 & 99.9968 & [99.982, 99.999] \\
BigGAN & 3,108 & 1 & 99.9678 & [99.818, 99.994] \\
DALLE & 12,897 & 2 & 99.9845 & [99.943, 99.996] \\
DDIM & 13,143 & 8 & 99.9391 & [99.880, 99.969] \\
DDPM & 15,313 & 22 & 99.8563 & [99.783, 99.905] \\
DF-GAN & 38,396 & 0 & 100.0000 & [99.990, 100.000] \\
GALIP & 32,529 & 0 & 100.0000 & [99.988, 100.000] \\
GigaGAN & 5,492 & 0 & 100.0000 & [99.930, 100.000] \\
Imagen & 9,487 & 0 & 100.0000 & [99.960, 100.000] \\
MAE & 1,678 & 0 & 100.0000 & [99.772, 100.000] \\
MAGE & 20,000 & 0 & 100.0000 & [99.981, 100.000] \\
Midjourney & 87,682 & 197 & 99.7753 & [99.742, 99.805] \\
SD & 177,068 & 158 & 99.9108 & [99.896, 99.924] \\
VQDM & 30,695 & 0 & 100.0000 & [99.987, 100.000] \\
VQGAN & 2,800 & 0 & 100.0000 & [99.863, 100.000] \\
VQVAE & 11,000 & 3 & 99.9727 & [99.920, 99.991] \\
starGAN & 3,089 & 0 & 100.0000 & [99.876, 100.000] \\
styleGAN & 16,000 & 38 & 99.7625 & [99.674, 99.827] \\
\bottomrule
\end{tabular*}
\end{minipage}
\end{center}
\medskip

\begin{center}
\begin{minipage}{\textwidth}
\centering
\captionsetup{type=table,hypcap=false}
\captionof{table}{WildFake real-source results recovered from the documented hierarchy in saved record identifiers. Each row is real-only, so only TNR and false positives are defined.}
\label{tab:WildFake_real_sources}
\small
\setlength{\tabcolsep}{3.5pt}
\begin{tabular*}{\textwidth}{@{\extracolsep{\fill}}lrrrr}
\toprule
Source & $N$ & FP & TNR & 95\% CI \\
\midrule
afhq & 6,387 & 15 & 99.7651 & [99.613, 99.858] \\
celebahq & 6,000 & 12 & 99.8000 & [99.651, 99.886] \\
church & 16,671 & 1 & 99.9940 & [99.966, 99.999] \\
coco & 32,770 & 7 & 99.9786 & [99.956, 99.990] \\
ffhq & 14,000 & 10 & 99.9286 & [99.869, 99.961] \\
imagenet & 19,358 & 11 & 99.9432 & [99.898, 99.968] \\
laion5b & 54,367 & 56 & 99.8970 & [99.866, 99.921] \\
\bottomrule
\end{tabular*}
\end{minipage}
\end{center}
\medskip

Finally, Table~\ref{tab:GenImage_subsets} retains all eight GenImage generator folders. Together, these legacy tables preserve the full record-level evidence that underlies the new comparison tables, confidence distributions, and curves.
\begin{center}
\begin{minipage}{\textwidth}
\centering
\captionsetup{type=table,hypcap=false}
\captionof{table}{GenImage: complete available subset statistics. Rates and AP are percentages; a dash denotes a metric undefined for a single-class subset. All counts are scored records.}
\label{tab:GenImage_subsets}
\small
\setlength{\tabcolsep}{3.5pt}
\begin{tabular*}{\textwidth}{@{\extracolsep{\fill}}lrrrrrrr}
\toprule
Subset & $N$ & Acc. (\%) & AP & TNR & TPR & FP & FN \\
\midrule
ADM & 12,000 & 99.9833 & 100.0000 & 99.9667 & 100.0000 & 2 & 0 \\
BigGAN & 12,000 & 99.9917 & 100.0000 & 99.9833 & 100.0000 & 1 & 0 \\
Midjourney & 12,000 & 99.9583 & 100.0000 & 99.9833 & 99.9333 & 1 & 4 \\
VQDM & 12,000 & 100.0000 & 100.0000 & 100.0000 & 100.0000 & 0 & 0 \\
glide & 12,000 & 100.0000 & 100.0000 & 100.0000 & 100.0000 & 0 & 0 \\
stable\_diffusion\_v\_1\_4 & 12,000 & 100.0000 & 100.0000 & 100.0000 & 100.0000 & 0 & 0 \\
stable\_diffusion\_v\_1\_5 & 16,000 & 99.9938 & 100.0000 & 99.9875 & 100.0000 & 1 & 0 \\
wukong & 12,000 & 99.9833 & 100.0000 & 99.9667 & 100.0000 & 2 & 0 \\
\bottomrule
\end{tabular*}
\end{minipage}
\end{center}
\medskip

\section{Real and Generated Confidence Distributions}

\begin{figure}[H]
\centering
\begin{minipage}{0.24\textwidth}\centering
\includegraphics[width=\linewidth]{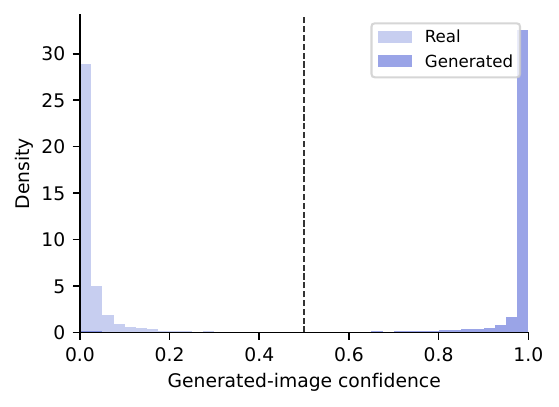}\\[-0.4em]\scriptsize Chameleon
\end{minipage}\hfill
\begin{minipage}{0.24\textwidth}\centering
\includegraphics[width=\linewidth]{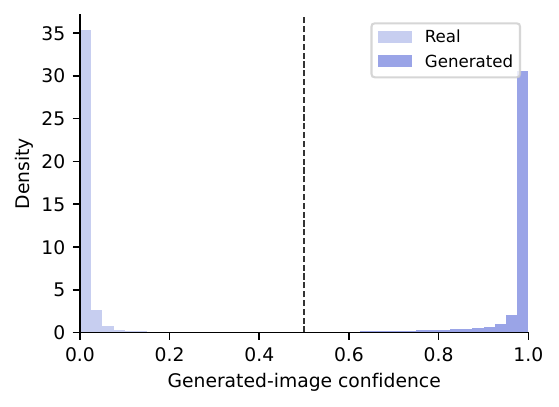}\\[-0.4em]\scriptsize OpenFake
\end{minipage}\hfill
\begin{minipage}{0.24\textwidth}\centering
\includegraphics[width=\linewidth]{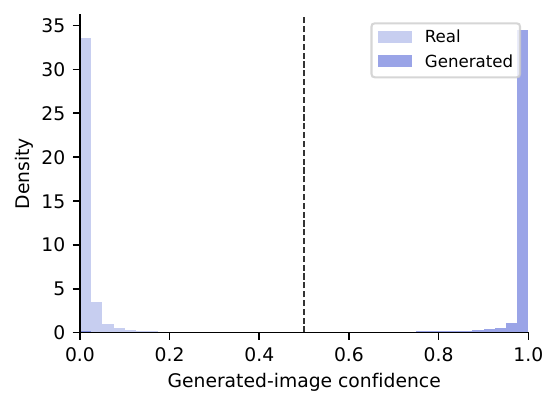}\\[-0.4em]\scriptsize Community Forensics
\end{minipage}\hfill
\begin{minipage}{0.24\textwidth}\centering
\includegraphics[width=\linewidth]{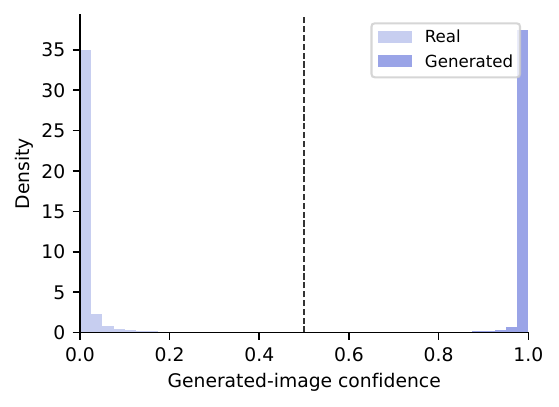}\\[-0.4em]\scriptsize AIGIBench
\end{minipage}
\vspace{0.4em}
\begin{minipage}{0.24\textwidth}\centering
\includegraphics[width=\linewidth]{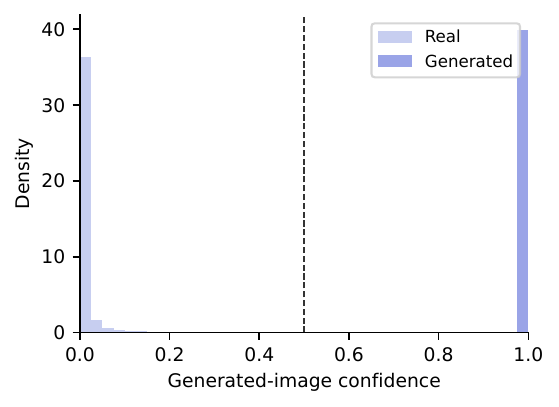}\\[-0.4em]\scriptsize AIGCDetectBenchmark
\end{minipage}\hfill
\begin{minipage}{0.24\textwidth}\centering
\includegraphics[width=\linewidth]{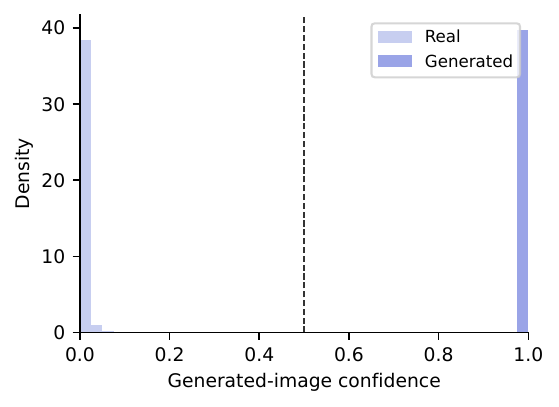}\\[-0.4em]\scriptsize WildFake
\end{minipage}\hfill
\begin{minipage}{0.24\textwidth}\centering
\includegraphics[width=\linewidth]{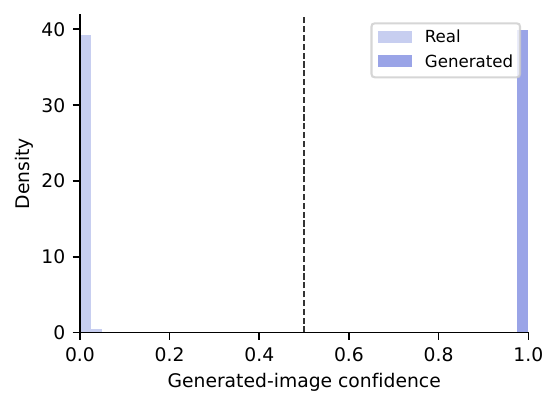}\\[-0.4em]\scriptsize GenImage
\end{minipage}
\caption{Neural Defend ARCAS 1B real-versus-generated confidence distributions. OpenFake is pooled across its still-image splits; WildFake is available-case.}
\label{fig:confidence_distributions}
\end{figure}

\begin{flushleft}
Figure~\ref{fig:confidence_distributions} shows the score distributions for real and generated images across the seven evaluated benchmark populations. The figure uses the same released benchmark populations cited in the corresponding Results subsections: Chameleon \cite{yan2025sanity}, OpenFake \cite{livernoche2025openfake}, Community Forensics \cite{park2025community}, AIGIBench \cite{li2025aigibench}, AIGCDetectBenchmark \cite{zhong2024patchcraft}, WildFake \cite{hong2025wildfake}, and GenImage \cite{zhu2023genimage}.
\end{flushleft}

\begin{flushleft}
The separation between the real and generated score distributions provides context for the reported precision, recall, APCER, and BPCER values. Where the two distributions overlap more strongly, the fixed threshold produces a larger class-specific error rate; where they are well separated, the thresholded confusion matrix and threshold-free ranking measures align more closely. OpenFake is pooled across still-image splits and WildFake is available-case, so their distributions retain the same scope qualifiers used in the main Results section.
\end{flushleft}

\section{Precision--Recall and DET Curves}

\begin{figure}[H]
\centering
\includegraphics[height=0.32\textheight,keepaspectratio]{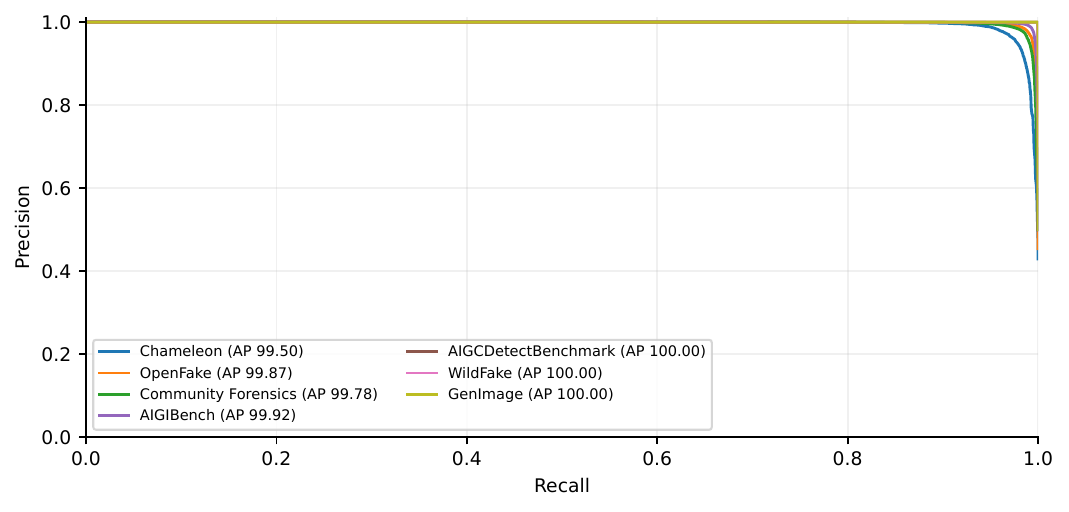}
\caption{Neural Defend ARCAS 1B precision--recall curves across the evaluated benchmark populations cited in the corresponding Results subsections.}
\label{fig:pr_curves}
\end{figure}

\begin{flushleft}
Figure~\ref{fig:pr_curves} compares precision--recall behavior across the benchmark populations. Precision--recall curves are particularly informative for assessing generated-image detection when class balance differs across releases, because they retain the trade-off between positive predictive value and generated-image recall over all score thresholds.
\end{flushleft}

\begin{flushleft}
The curves remain near the high-precision, high-recall region for most populations, but their detailed shapes identify where a fixed 0.5 threshold may not be the most favorable operating point. This is why the paper reports both threshold-free AP/AUROC and fixed-threshold error measures instead of treating them as interchangeable.
\end{flushleft}

\begin{figure}[H]
\centering
\includegraphics[height=0.32\textheight,keepaspectratio]{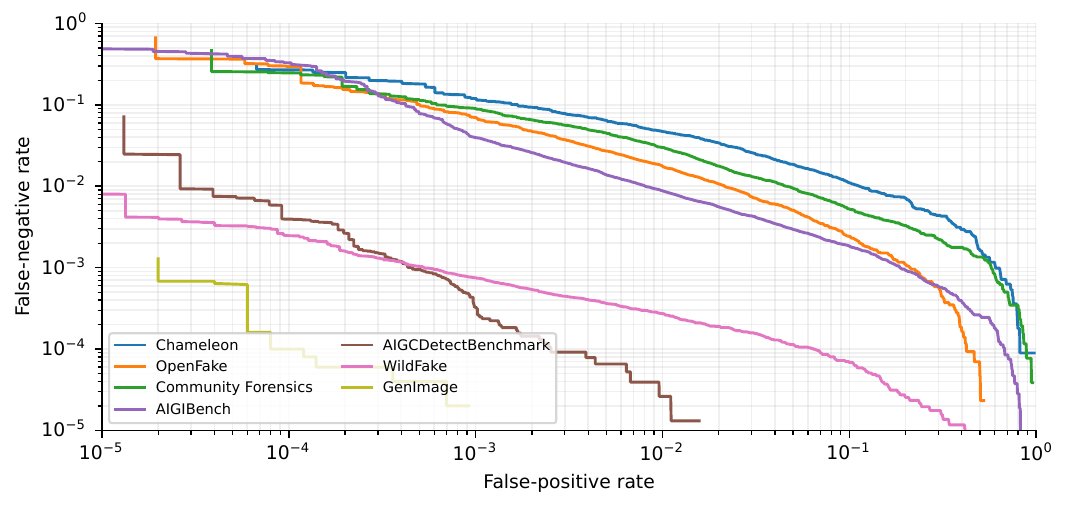}
\caption{Neural Defend ARCAS 1B detection-error tradeoff (DET) curves across the evaluated benchmark populations cited in the corresponding Results subsections.}
\label{fig:det_curves}
\end{figure}

\begin{flushleft}
Figure~\ref{fig:det_curves} presents the same score populations as detection-error tradeoff curves. DET coordinates make the false-positive and false-negative trade-off visible on scales that reveal low-rate differences which can be compressed in ordinary ROC-style plots.
\end{flushleft}

\begin{flushleft}
The DET curves support the main-text conclusion that residual errors are not identical across datasets. Benchmarks with more generated-image misses have a different low-error trade-off from AIGCDetectBenchmark, where false positives make up most of the observed residual. The separate low-FPR ROC figure remains in the main cross-benchmark analysis because it directly supports that section's deployment-oriented discussion.
\end{flushleft}

\end{document}